\documentclass[
amsmath, 
amssymb, 
aps, 
pra,
twocolumn,
floatfix,
nofootinbib
]{revtex4}

\usepackage[utf8]{inputenc}
\usepackage{titlesec}
\usepackage[colorlinks]{hyperref}
\usepackage{amsmath,amssymb}
\usepackage{bbold}
\usepackage{float}
\usepackage{dcolumn}
\usepackage{bm}
\usepackage{dsfont}
\usepackage{url}
\usepackage{mathtools}
\usepackage{amsfonts}
\usepackage{braket}
\usepackage{cancel}
\usepackage{bbold}
\usepackage{makecell}
\usepackage[normalem]{ulem}
\usepackage{xcolor}
\usepackage{comment}

\begin{document}
\title{Improving the Performance of Quantum Cryptography\\by Using the Encryption of the Error Correction Data}
\author{V. A. Pastushenko}
\author{D. A. Kronberg}
\address{Terra Quantum AG, Kornhausstrasse 25, 9000 St. Gallen, Switzerland
\\
Correspondense should be sent to dk@terraquantum.swiss (D.A.K.)}

\begin{abstract}
Security of quantum key distribution (QKD) protocols relies solely on quantum physics laws, namely, on the impossibility to distinguish between non-orthogonal quantum states with absolute certainty.
Due to this, a potential eavesdropper cannot extract full information from the states stored in their quantum memory after an attack despite knowing all the information disclosed during classical post-processing stages of QKD.
Here, we introduce the idea of encrypting classical communication related to error-correction in order to decrease the amount of information available to the eavesdropper and hence improve the performance of quantum key distribution protocols.
We analyze the applicability of the method in the context of additional assumptions concerning the eavesdropper's quantum memory coherence time and discuss the similarity of our proposition and the quantum data locking (QDL) technique.
\end{abstract}
     
\maketitle

\section{Introduction}

The main goal of quantum key distribution (QKD)~\cite{Gisin2002,Pirandola2020,QKD_industrialization} is to generate a secret key between two remote users (Alice and Bob), with the security of the key not based on computational assumptions on a potential eavesdropper (Eve).
The first ever QKD protocol, BB84, was proposed by Bennett and Brassard~{\cite{BB84}}.
In the protocol, the legitimate sender, Alice, encodes random bit string into polarization states of single photons and sends them to the legitimate receiver, Bob.
For the encoding purposes, Alice randomly chooses one of two orthogonal polarization bases,  while all the four states form a non-orthogonal set. Bob uses a random basis guess when conducting his measurement. 
The bit values corresponding to wrong guesses on Bob's side are discarded after a round of classical communication, which provides an advantage to the legitimate users.

The security of quantum key distribution is based on the indistiguishability of the states available to the eavesdropper. 
For QKD protocols, any eavesdropping attempt leads to a disturbance of the states at the receiver side, and the value of the disturbance in the observed parameters allows the legitimate users to estimate the quantum states of Eve and hence bound her information. 
When this information is below the information available to the legitimate users, they can use classical post-processing methods to distill a secret key.

Along with QKD, an adjacent technology of quantum data locking (QDL) is of interest. 
In the QDL scenario, a potential adversary does not have enough data to perform a correct measurement of the available quantum states, while the total extractable information may be relatively high. 
Hence, a relatively short amount of classical data can lock a large amount of information.
The first QDL protocol was proposed in~{\cite{QDL_first_ever_2004}} and resembles BB84, as the same two bases are used. Now, there is a single pre-shared bit that specifies the basis choice for each bit of Alice's string and for Bob's measurement at every position.
As was shown in~{\cite{QDL_first_ever_2004}}, the eavesdropper who has intercepted $N$ signals obtains no more than $N/2$ bits of information; hence, the single secret bit is sufficient for ``locking'' $N/2$ bits of classical information.
Later, other QDL protocols based on different principles were introduced~{\cite{QDL_Boixo, QDL_Lupo_2014}}.

Here, we propose a method that can increase the secret key rate in QKD by simple additional actions of the legitimate users, namely the encryption of information that is disclosed during error correction. 
Encryption of the postprocessing data was also used in~\cite{koashi2003secure} to simplify the security proof, but here, we discuss its usage for a different purpose---for a QDL-like technique that does not allow an eavesdropper to perform the best possible measurement. 

The paper is organized as follows. In Section~\ref{sec:Background}, we recall the main stages of prepare-and-measure QKD protocols and discuss various types of eavesdropping attacks. 
Section~\ref{sec:main_idea} is devoted to the condition sufficient for quantum accessible information additivity and its application to QKD in a QDL scenario.
Section~\ref{sec:QKD_application} addresses the case of the most general eavesdropping attacks and the limitations of the method we propose.
Finally, we discuss the results in Section~\ref{sec:discussion}.

\section{Background}
\label{sec:Background}
The detailed description of the stages for the typical QKD protocol can be found in reviews, see, e.g.,~\cite{Gisin2002,Pirandola2020}.
Here, we focus on the four stages that are the most significant for our~study:
\begin{enumerate}
    \item The quantum states are sent via a quantum channel from Alice to Bob, with Bob performing an appropriate measurement.
    Then, the legitimate users utilize a classical authenticated channel to perform key sifting and/or basis reconciliation. 
    After this procedure, Alice and Bob have correlated but not yet coinciding classical bit sequences also correlated with Eve.
    \item The legitimate users estimate the intervention of Eve and the information available to her based on the data observed at the receiver's side. 
    The most significant parameter is quantum bit error rate (QBER), but other parameters including the visibility, attenuation, or gain of different classes of states can be utilized as well~\cite{Decoy1,Decoy2,COW}. 
    The aforementioned estimate can be performed by disclosing a part of the signals, which is then removed from the key as it is not secret any more.
    \item The legitimate users perform error correction, which provides them with coinciding keys correlated with Eve. 
    Classical error correction codes~\cite{kiktenko2017symmetric} or the Cascade method~\cite{brassard1994cascade} may be used. 
    The legitimate users take into account that some secret data are disclosed during error correction.
    The disclosed data actually specify the set of codewords used by Alice, 
    e.g., for linear codes, the syndrome specifies that ``a string of Alice is the one that produces the following syndrome'', while the check matrix of the linear code may be fixed for many communication sessions. 
    The Cascade method uses the interactive exchange of parity bits, which also specify the set of possible bit sequences used by Alice.
    \item Finally, the privacy amplification stage follows.
    This results in a shorter key with very low correlation with the eavesdropper. 
    The length of the final key depends on the data observed by the legitimate users and, correspondingly, their estimate of Eve's and their own information. 
    In addition, the security proof, i.e., the proof of the statement that the key obtained with this formula is secure according to the security parameter (see, e.g.,~\cite{security_parameter_Trushechkin, BB84_security_parameter}), is the main theoretical element for the QKD protocol.
\end{enumerate}

The classical or quantum data available to the participants after each of these stages can be described by a quantum states in the joint Hilbert space of Alice--Bob--Eve, with classical states of Alice and Bob being diagonal density matrices in some fixed basis. 
The total state depends on the attack performed by the eavesdropper.

All the eavesdropping attacks on QKD protocols can be categorized into three nesting groups.
The most general type of attack entails conducting a joint unitary transformation with ancilla on an arbitrary number of signal quantum states and subsequent collective measurement of the ancillary system. 
This sequence of actions conducted by an eavesdropper is called a coherent attack.
If the eavesdropper is limited to conducting an individual unitary transformation with an ancilla of every signal state followed by a collective measurement of all the ancillary systems, then their intervention can be attributed to a narrower class of collective attacks.
Finally, each attack including only individual unitary transformations and individual measurements of ancillary systems belongs to the subset of individual~attacks.

Full security analysis of any QKD protocol, i.e., proving its unconditional security, requires considering Eve to be able to perform any action allowed by the laws of quantum physics, i.e., operating in the set of coherent attacks.
However, on the basis of the quantum version of the de Finetti representation theorem~\cite{de_Finetti1,de_Finetti2}, it was shown that coherent attacks are not more powerful than collective ones and, thus, the considered set can be narrowed: the only condition required for the statement to be true is that a given QKD protocol is permutation-invariant, i.e., invariant under arbitrary permutations of quantum channel uses~\cite{Renner_phd}.
The condition can be satisfied for the majority of standard QKD protocols including BB84\,\cite{BB84}, six-state\,\cite{Six_State}, and B92\,\cite{B92} by introducing an additional step to their structure: legitimate users should publicly agree on a random permutation of raw key bits right after the first stage~\cite{Renner_phd, Renner_Gisin_Kraus}.

The secret key generation rate of a QKD protocol is defined as the maximum speed (per bit) at which a secret key can be distributed---here, secret means that the eavesdropper's knowledge about it is asymptotically small.
In the case of classical key distribution protocols, the secret key rate can be calculated according to the Sciszar and Korner's equation~\cite{Sciszar_Korner}:
 
\begin{equation}
    \text{rate}=I\left(\text{X},\text{Y}\right)-I(\text{X},\text{Z}),
    \label{classical_key_rate}
\end{equation} 

where $I$ is mutual information between two classical systems (X, Y, and Z stand for the random variable describing Alice's, Bob's, and Eve's systems, respectively): $I(\text{X}_1,\text{X}_2)=H(\text{X}_1)+H(\text{X}_2)-H(\text{X}_1\text{X}_2)$, with $H$ being the Shannon entropy of a random variable.
The legitimate users are to estimate the range of attacks that can be feasibly conducted by Eve (using the assumptions concerning her computational powers) and use the value of mutual information $I(\text{X},\text{Z})$ for the most effective one.
The expression (\ref{classical_key_rate}) should be implemented in the case of direct reconciliation---when Alice's bit string is considered to be correct and Bob has to amend his string. 
Although, the equation can be easily modified for the case of reverse reconciliation, which corresponds to Alice and Bob changing roles and thus leads to switching X and Y in the equation.

Transitioning into the quantum cryptography framework implies that legitimate users no longer use any assumptions related to the eavesdropper's computational powers; they rely only on the laws of quantum mechanics in order to determine the range of attacks that could have been conducted.
They have to determine the set $\Gamma$ of all quantum states $\rho_{\text{AB}}$ that can be shared between them according to the set of observed data.
For each state $\rho_{\text{AB}}$ describing the system shared between Alice and Bob, $\rho_{\text{ABE}}$ is defined as its arbitrary purification and includes Eve.
Then, after the measurements conducted on Bob's and Alice's ends combined with the reconciliation procedure, the final state $\rho_{\text{XYE}}$ describes the system shared between Alice, Bob, and Eve right after stage 1 (conditioned on the conclusive result, i.e., when the position survived key sifting):
 \begin{equation*}
    \rho_\text{XYE}=
    \sum\limits_{x,y}
    p_{xy}\ket{x}\bra{x}_\text{A}
    \otimes\ket{y}\bra{y}_\text{B}
    \otimes\rho_\text{E}^{xy}.
\end{equation*} 
where $p_{xy}$ is the joint probability of Alice sending classical value $x$ and Bob obtaining the result $y$; $\ket{x}_{\text{A}}$ and $\ket{y}_\text{B}$ denote the classical states of the legitimate user's systems corresponding to the values. Then, Eve's ensemble $\mathcal E_\text{E} = \{(p_x, \rho_x)\}_x$ of quantum states $\rho_x$ corresponding to different bit values on Alice's side reads
 \begin{equation*}
    \mathcal E_\text{E}=\left\{\left(p_{x}\equiv\sum\limits_{y}p_{xy},\;\rho_{x}\equiv\sum\limits_{y}p_{xy}\cdot\rho_\text{E}^{xy}\right)\right\}_{x}.
\end{equation*} 
This knowledge is sufficient to upper bound the information available to Eve.
The Holevo bound~\cite{holevo1973bound} can be used for the purpose, as the Holevo quantity $\chi(\mathcal E_\text{E})=S\left(\sum_x p_x\rho_x\right)-\sum_x p_x S\left(\rho_x\right)$, where $S(\rho) = {\rm Tr}\rho\log_2\rho$ is the von Neumann entropy, upper bounds the accessible information

\begin{equation*}
    I_\text{acc}(\mathcal E_\text{E}) = \sup\limits_{\mathcal M_{\text{Z}\leftarrow\text{E}}}
    I(\text{X},\text{Z}),
\end{equation*}

which can be extracted from the ensemble $\mathcal E_\text{E}$ by performing the most optimal of all the quantum measurements $\mathcal M_{\text{Z}\leftarrow\text{E}}$ on the system E.
The estimation allows transitioning from the classical Equation (\ref{classical_key_rate}) to the equation lower-bounding the secret key rate in QKD:
 \begin{equation}
    \text{rate}\geq\inf\limits_{\rho_{\text{AB}}\in\Gamma}\Big(I\left(\text{X}:\text{Y}\right)-\chi\left(\mathcal E_\text{E}\right)\Big),
    \label{key_rate_Holevo}
\end{equation} 
which is the content for the seminal Devetak--Winter result~\cite{Devetak_Winter}. Here, we described the intuition behind this result based on  Sciszar and Korner classical equation, while a complete proof of (\ref{key_rate_Holevo}) is much more complex.

\section{The Method Description}
\label{sec:main_idea}

We use Theorem 2 in~\cite{Sasaki_Kato}, bounding the accessible information in new conditions, which we want to achieve in quantum cryptography by simple actions of the legitimate users.

Let us briefly describe this result of~\cite{Sasaki_Kato}. 
The above-mentioned theorem provides a sufficient condition for the additivity of accessible information, which is the independent use of all the states' combinations.
To put it in formal terms: if a multipartite ensemble of quantum states $\mathcal E^N=\left\{\xi^N_i,\rho^N_i\right\}_i$ has a product form, i.e., if $\xi_i^N=\xi_{i_i}^{\text{\small{(1)}}}\cdot\ldots\cdot\xi_{i_N}^{\text{\small{($N$)}}}$ and $\rho_i^N=\rho_{i_1}^{\text{\small{(1)}}}\otimes\ldots\otimes\rho_{i_N}^{\text{\small{($N$)}}}$, the quantum accessible information of the ensemble is additive: 
 \begin{equation}
    I_\text{acc}\left(\mathcal E^N\right)
    =
    I_\text{acc}\left(\mathcal E^{(1)}\right)+\ldots+I_\text{acc}\left(\mathcal E^{(N)}\right),
    \label{additivity}
\end{equation} 
where $\mathcal E^{\text{\small{$(n)$}}}=\big\{\big(\xi^{\text{\small{$(n)$}}}_{i_n},\rho^{\text{\small{$(n)$}}}_{i_n}\big)\big\}_{i_n}$ is the $n$th partial ensemble describing the $n$th system.
Thus, for such product-form ensembles, collective measurements do not provide any advantage over a sequence of independent individual measurements in terms of extracted information.

If a given QKD protocol is permutation-invariant, the set of considered eavesdropping attacks can be narrowed to collective ones.
Thus, after $N$ channel uses, Eve's ensemble $\mathcal{E}_E^N$ satisfies the conditions of this theorem: the states of the ensemble have product form,
as well as the states' probabilities, which are distributed according to the initial probability distribution on the Alice side, as Alice sends the states independently in each position. 
Hence, if Eve performs the measurement at this time, the mutual information between the result of Eve's measurement (contained in a classical system E) and the classical value sent by Alice (system X) is bounded by additive accessible information: 

\begin{equation}
    \label{Eve_additive_bound}
    I^N(\text{X}, \text{E}) \leq
    I_\text{acc}\left(\mathcal{E}_\text{E}^N\right)
    = N I_\text{acc}\left(\mathcal E_\text{E}\right).
\end{equation}

Now, observe that when Alice and Bob perform the error correction step, they change the probability distribution, as they disclose the set of possible codewords, and the new probabilities do not have the product form. 
Hence, the estimate (\ref{Eve_additive_bound}) do not hold any longer, and Eve's information may overcome $N I_\text{acc}\left(\mathcal E_\text{E}\right)$. 
This is the subject of the quantum coding theorem\,\cite{HSW_H,HSW_SW}: if the sender and the receiver have fixed the set of the codewords, then the receiver may perform a collective measurement which allows the Holevo capacity to be achieved.
The result of~\cite{Sasaki_Kato} therefore states that without coding (i.e., without a non-trivial subset of all the possible bit strings to be the codewords), the users cannot achieve any superadditive information, let alone the Holevo capacity.
Within our framework, this means that Eve, who plays the role of the receiver now, does not get the amount of information characterized by the Holevo quantity and is limited by a more strict bound. Hence, using the Holevo capacity as the estimate for Eve's information becomes too pessimistic.

Disclosing additional information may be regarded as implementing the QDL protocol between Alice and Eve, who are now in the conditions of quantum coding theorem. Here, as it happens in QDL, Eve cannot perform the proper measurement without additional information but can do so after obtaining it, namely, after knowing the set of codewords to perform a collective measurement (see Section 4 in~\cite{QDL_Boixo}).

Our idea is that the legitimate users should not change the probabilities of the states available to the eavesdropper. 
They can avoid doing this by encrypting the information disclosed during error correction. 
When Eve gets no additional information, she is restricted by (\ref{additivity}), and her information obtained with the best possible measurement is still below $N I_\text{acc}\left(\mathcal E_\text{E}\right)$. 

A potential problem may appear due to the information disclosure taking place during privacy amplification procedure, since it makes Eve's states statistically dependent, and thus her ensemble $\mathcal E_\text{E}^N$ loses product form---see Section~\ref{sec:QKD_application} for detail.
However, in the case when Eve is forced to measure the obtained quantum states before the legitimate users begin privacy amplification routine, the method works well.
Instead of disclosing the $H(\text{X}|\text{Y})$ bits during the error correction stage, the legitimate users would consume a part of the pre-distributed key in order to encrypt the classical communication using the one-time pad.
Here, $H(\text{X}|\text{Y})=H(\text{XY})-H(\text{Y})$ is the conditional entropy, which characterizes the lack of knowledge about X when the full information about Y is provided~\cite{Cover_Thomas}.
At the same time, after Eve's measurement, when all the participants operate with classical data, the legitimate users are able to substitute the value $\chi(\mathcal E_\text{E})$ in the Devetak--Winter equation with $I_\text{acc}(\mathcal E_\text{E})$, thus obtaining a higher key generation rate without compromising the security of the whole scheme:
 \begin{equation}
    \text{rate}\geq\inf\limits_{\rho_{\text{AB}}\in\Gamma}\Big(I\left(\text{X}:\text{Y}\right)-I_\text{acc}\left(\mathcal E_\text{E}\right)\Big).
    \label{acc_info_rate}
\end{equation} 
Recall that the set $\Gamma$ includes all the bipartite states that can be shared between the legitimate users based on the statistics of their measurement results.
Let us emphasize that no hardware modification is required for this secret key rate boost.

In order to force Eve to measure her states at an early stage and use the bound (\ref{Eve_additive_bound}), legitimate users can employ some additional assumptions concerning Eve's technical abilities.
The assumption about the upper-bound on the eavesdropper's quantum memory decoherence time is a natural one typically utilized in a quantum data locking scenario as well as in a QKD scenario with a restricted Eve.
This allows us to benefit from postponing the privacy amplification for an amount of time sufficient for the eavesdropper's quantum memory to lose coherence or from encrypting all the classical communication necessary for the stage with an asymmetrical cipher such as AES. 
In the latter case, the legitimate users are to assume that Eve cannot break a chosen encryption during her quantum memory coherence time.
The tactic allows legitimate users to assume that an eavesdropper is to conduct the measurement without any additional knowledge associated with the information from privacy amplification.

In this scenario, the size of Eve's quantum memory is not limited, and her ability to conduct collective measurements is not restricted as well---this significantly distinguishes the approach we propose from the bounded quantum storage model (BQSM), which is built on the assumption concerning the maximal number of quantum states that an eavesdropper can keep in their quantum memory~\cite{BQSM_Pironio, BQSM_Damgard, Pirandola2020}.
Nevertheless, our approach makes collective attacks no more efficient than individual ones and thus eliminates the necessity to consider any eavesdropping relying on quantum memory capable of storing more than one quantum state at a time.
Moreover, Eve may know all the information concerning bit reconciliation and post-selection procedures, as the availability of the data does not destroy the statistical independence of separate signal states.

Thus, we propose a modification of the initial scheme presented in Section~\ref{sec:Background}: the first two stages may remain unchanged, while the subsequent stages are modified in the following way:
\begin{enumerate}
    \item[3'.] Alice and Bob perform error correction in a standard manner, with the only difference that now they utilize a private channel for the purpose, i.e., all the communication conducted at this stage is encoded by one-time pad cipher using the pre-distributed key.
    Thus, they deprive Eve of any information concerning codewords choice.
    \item[4'.] The legitimate users perform privacy amplification with some delay sufficient for Eve's quantum memory to lose coherence or encrypt all the communication necessary for the privacy amplification stage (in contrast to the previous step, an asymmetrical cipher such as AES is to be utilized). 
    The compression ratio depends on the legitimate users' assumptions concerning Eve.
    If the decoherence time of her quantum memory is considered to be limited by some finite value, then privacy amplification goes according to Equation (\ref{acc_info_rate}) up to a minor value of the extra key needed for symmetric encryption.
\end{enumerate}

Notably, the method relies on using a pre-distributed secret key for encoding a part of classical communication. 
However, this does not change the common QKD paradigm, as any quantum key distribution protocols begin with an authenticating classical channel using a relatively short initial key (for this reason, key distribution protocols have an alternative name: ``key expansion protocols'').
Our approach leads to the necessity of a longer initial key for the very first round of key distribution, while no data on the raw key are disclosed during the error correction stage, in contrast with the conventional scenario.
The key for encoding classical communication in each subsequent round is to be taken from the secret string distributed in the preceding one.

The scheme works well in an asymptotic case, when the size of the distributed key is large enough and post-processing procedures are asymptotically efficient.
However, in practice, the difficulties related to the finiteness of the key length lead us to the paradigm of $\varepsilon$-secure data exchange~\cite{security_parameter_Trushechkin,Portman_Renner}.
Additional difficulties appear when a part of the generated secret key is utilized in the following round of communication, resulting in the overall security slightly degrading with the number of rounds.
It worth noting that within our framework, the security level decreases more quickly than in conventional QKD schemes, since we propose using larger amounts of the previously distributed key for the next round.
Thus, an accurate analysis of our method beyond the asymptotic case is a perspective and important area for future research.

In summary, the modified scheme involves encrypting classical communication (during error correction and privacy amplification stages) and leaving a part of generated key for the next round.
Combined with the assumption concerning the upper bound on the decoherence time of Eve's quantum memory, this allows the legitimate users to come to classical signals analysis and the equation analogous to the result of Sciszar and Korner (\ref{classical_key_rate}), where Eve's information is bounded according to (\ref{acc_info_rate}) operating with restricted accessible information (\ref{Eve_additive_bound}).

\section{Beyond Memory-Restricted Scenario}\label{sec:QKD_application}

If Eve is not forced to conduct her measurements right after the error-correction stage, it is more beneficial for her to measure the states later---when she will be able to take into account the information disclosed during the privacy amplification procedure. 
In this case, an observable that was optimal when measuring the original states can become non-optimal for measuring the states after information processing.
In \,\cite{on_classical_data_processing}, an explicit example was provided, which shows that the strategy yields gain for Eve, i.e., that classical processing of states of a quantum ensemble changes the set of observables providing accessible information. 

The example is based on considering a quantum ensemble $\mathcal E_{\text{init}}$ obtained as the result of a simple two-letter classical-quantum channel utilized twice (the lower index ``init'' indicated that the ensemble is obtained before the classical information processing).
 \begin{multline*}
    \mathcal E_\text{init}=
    \Bigg\{
    \left(\frac{1}{4},\;\sigma_{0}\otimes\sigma_{0}\right),\  \left(\frac{1}{4},\;\sigma_{0}\otimes\sigma_{1}\right),\\
    \left(\frac{1}{4},\;\sigma_{1}\otimes\sigma_{0}\right),\  \left(\frac{1}{4},\;\sigma_{1}\otimes\sigma_{1}\right)\Bigg\},
\end{multline*} 
where the equiprobable letter states (described by density operators on two-dimensional Hilbert space $\mathcal{H}$) $\sigma_{0}$ and $\sigma_{1}$ are pure and can be represented as real vectors in some orthonormal basis $\left\{\ket{0},\ket{1}\right\}\subset \mathcal H$:
\begin{gather*}
    \forall x\in\{0,1\}:\;\sigma_x=\ket{\psi_x}\bra{\psi_x},\\|\psi_x\rangle =\cos\alpha|0\rangle +(-1)^x\sin\alpha|1\rangle.
\end{gather*} 
According to~\cite{Sasaki_Kato}, an optimal strategy for extracting the maximal amount of information from the quantum ensemble $\mathcal E_{\text{init}}$ consists in conducting two independent local measurements (measurements in the Hadamard basis).
Then, a simple classical data processing corresponding to an XOR operation can be considered.
It merges some states and transforms $\mathcal E_\text{init}$ into an ensemble 
\begin{multline*}
    \mathcal E
    =
    \Bigg\{\left(\frac{1}{2},\:\frac{1}{2}\sigma_0\otimes\sigma_0+\frac{1}{2}\sigma_1\otimes\sigma_1\right),
    \\
    \left(\frac{1}{2},\:\frac{1}{2}\sigma_0\otimes\sigma_1+\frac{1}{2}\sigma_1\otimes\sigma_0\right)\Bigg\}.
\end{multline*} 
It was shown in~\cite{on_classical_data_processing} that there exists such a range of $\alpha$ values for which it is true that any observable providing $I_\text{acc}\left(\mathcal E\right)$ has to include entangled operators.
Moreover, the measurement in the Bell basis is always the optimal measurement strategy for $\mathcal E$.
Thus, classical information processing can significantly change the structure of the optimal observable.
However, the question of the existence of classical data processing operations preserving an optimal observable remains, to our knowledge, open.

Privacy amplification in QKD is an important special case of classical data processing.
In particular, the considered XOR operation can be an element of some universal hash functions family used for privacy amplification.
This explains the significance of the example in the context of our study: it proves that there exist privacy amplification procedures turning the disclosure of privacy amplification-related information into QDL-type communication between legitimate users and an eavesdropper.

Notably, if the opposite statement was true and any observable that was optimal before classical data processing remained optimal after the operation, then it would not have been important whether an eavesdropper conducted their measurement before or after obtaining privacy amplification-related information (this would not influence the efficiency of their attack).
In this \textit{imaginary} situation, we could have  constructed a statement about our method's applicability while leaving privacy amplification data exchange completely unencrypted.

To our knowledge, the problem of determining an exact upper bound on the information available to Eve conducting her measurement after the privacy amplification stage remains open due to the difficulty of calculating the accessible information for an ensemble of states of a high-dimensional space~\cite{acc_info_difficult}.
At the moment, this fact limits the applicability of the proposed method in the case of no assumptions made about the eavesdropper's quantum memory storage time.
Nevertheless, future research may discover ways of calculating the value that are sufficiently easy to be practically implemented.
Currently, it is known that the above-mentioned value is upper bounded by the Holevo quantity and lower bounded by additive accessible information (which is much easier to calculate than the exact value of information available to Eve due to a significantly lower dimensionality of the problem)---in both cases, we are to take the influence of the privacy amplification into account, i.e., to subtract the corresponding number of bits as if we worked with classical data.

Note that in contrast to the case of error-correction data encryption, using the one-time pad for encrypting communication related to the privacy amplification procedure would not necessarily guarantee a gain in the secret key generation rate, as it consumes a relatively large additional amount of the pre-distributed key because of the large number of hash functions in the family, e.g., a large bit string is needed to specify the Toeplitz {\linebreak} matrix~\cite{kiktenko2016post}.

\section{Discussion}\label{sec:discussion}

In this paper, we proposed a method of increasing secret key distribution rates in the existing QKD protocols by encrypting classical communication or delaying it in the case of restrictions imposed on the eavesdropper's quantum memory coherence time.
Notably, it is universal (its applicability does depend on the specific protocol; despite the fact that in this work we consider only prepare-and-measure protocols, the method can be applied to entanglement-based QKD as well) and can be implemented just by modifying existing post-processing routines without introducing any changes to the hardware part of QKD~realization.

Under the assumption of limited coherence time of the eavesdropper's quantum memory, the method allows us to show that collective attacks become no more effective than individual ones.
If for a given QKD protocol coherent eavesdropping strategies have no advantage over collective, then individual attacks are the only ones to consider, and the key rate formula can be modified to operate with additive quantum accessible information.

Without any assumptions concerning the technical abilities of a potential eavesdropper, the key rate formula can be modified as well.
However, in such a case, the new bound for superadditive accessible information is still, to our knowledge, an open question.
Thus, we emphasize that the paper does not claim to provide a full security proof for QKD protocols in case of the method being implemented.

Note that the method inherits the disadvantages of quantum data locking: the disclosure of one bit of classical information that is meant to be secret (in this case, it is data related to error correction and privacy amplification procedures) may lead to an eavesdropper obtaining more than one bit of additional information.
This leads to increased demands on the safekeeping of the classical data.
Thus, the method does not provide composable security~\cite{Portman_Renner} against an eavesdropper who has access to unbounded quantum resources.
Nevertheless, the method provides everlasting security~\cite{renner_wolf} in a narrow sense: if an eavesdropper does not have access to quantum memory with storage time being sufficiently long at the moment of performing an attack (if the legitimate users have strong arguments in favor of this assumption), then no future advances in quantum memory can make an already distributed key less secure.

\bigskip
\section*{Acknowledgments}
D.A. Kronberg is grateful to E.O. Kiktenko, A.S. Trushechkin, and A.S. Holevo for useful discussions.

\end{document}